\documentclass{ws-ijmpd}
\usepackage[super,compress]{cite}
\usepackage[caption=false]{subfig}
\begin{document}

%%%%%%%%%%%%%%%%%%%%% Publisher's Area please ignore %%%%%%%%%%%%%%%
%
\catchline{}{}{}{}{}
%
%%%%%%%%%%%%%%%%%%%%%%%%%%%%%%%%%%%%%%%%%%%%%%%%%%%%%%%%%%%%%%%%%%%%

\title{BACK-REACTION IN RELATIVISTIC COSMOLOGY}

\author{Timothy Clifton}

\address{School of Physics and Astronomy,\\
  Queen Mary University of London, \\Mile End Road,
  London E1 4NS, UK.\\
t.clifton@qmul.ac.uk}

\maketitle

%\begin{history}
%\received{Day Month Year}
%\revised{Day Month Year}
%\end{history}

\begin{abstract}
We introduce the concept of back-reaction in relativistic cosmological modeling.  Roughly speaking, this can be thought of as the difference between the large-scale behaviour of an inhomogeneous cosmological solution of Einstein's equations, and a homogeneous and isotropic solution that is a best-fit to either the average of observables or dynamics in the inhomogeneous solution. This is sometimes paraphrased as `the effect that structure has of the large-scale evolution of the universe'. Various different approaches have been taken in the literature in order to try and understand back-reaction in cosmology. We provide a brief and critical summary of some of them, highlighting recent progress that has been made in each case. 
\end{abstract}


\ccode{PACS numbers: 98.80.Jk}

\section{What is Back-Reaction?}	

The term `back-reaction' is often used in cosmology to mean `the effect that structure has on the large-scale evolution of the universe, and observations made within it'. Implicit within this statement are a number of fundamental problems that have yet to be fully understood. These include:
\begin{itemize}
\item[1.] What is meant by the large-scale expansion of space in an inhomogeneous universe, and how should it be calculated?
\item[2.] How should we link the large-scale expansion of an inhomogeneous space-time with the observations made within it?
\item[3.] How can we create relativistic cosmological models sophisticated enough to investigate these problems?
\end{itemize}
Let us now briefly consider each of these points, before moving on to discuss recent attempts to understand them.

Point 1 above alludes to the fact that in relativistic theories what we mean by the spatial separation of any two astrophysical objects depends on how we choose to foliate the universe with hyper-surfaces of constant time. In a spatially homogeneous universe, or a universe with an irrotational matter content, natural-looking choices might present themselves. In general, however, we should be free to make any number of choices. This then presents a problem: If the distance between any two astrophysical objects is in general foliation dependent, and we have no preferred foliation, then how should we go about defining the rate of change of distance between objects, and hence the expansion of the universe? In the end, the answer to this question will depend on exactly what one is trying to achieve, and is complicated considerably by the fact that in cosmology one is often interested in non-local averages (a notoriously difficult concept to define in general relativity). Below we will consider several different cases of interest.

Point 2 is a subsequent problem that needs to be addressed, once a concept of `large-scale expansion' exists that one is prepared to consider. It is not in general the case that observations made in an inhomogeneous geometry will have a straightforward correspondence with the observables that one would measure in a spatially homogeneous and isotropic universe with the same rate of expansion on large scales. That is, even if one succeeds in finding a good description for the large-scale expansion of the universe, then one still needs to do further work in order to relate this to observations made in the underlying inhomogeneous space-time. Once again, this is complicated considerably by the fact that we are often interested in the average of observables. This is in general a highly non-trivial problem, and below we will review some recent progress towards understanding it.

Finally, point 3 is related to the fact that in order to test proposed solutions to the problems posed in points 1 and 2 it is of considerable interest to have cosmological models  that are sophisticated enough to allow at least some of the interesting behavior that we expect in general. This is an extremely difficult problem. Although many inhomogeneous cosmological solutions to Einstein's equations are known\cite{kra}, most of these solutions are restricted either because they are required to exhibit a high degree of symmetry, or because they are algebraically special. Constructions such as the ``Swiss cheese" models allow some potential progress to be made, but are themselves severely restricted by the boundary conditions at the edge of each ``hole". New approaches are required to make further progress in this area, and, once again, we will discuss some recent progress below.

In Section \ref{space} we consider approaches based on averaging over a set of prescribed spatial hyper-surfaces. In Section \ref{spacetime} we consider approaches based on averaging in four dimensions. Section \ref{models} contains a discussion of some models that may be of use for studying averaging, and in Section \ref{discussion} we provide a few closing comments.

\section{Spatial Averaging Approach}
\label{space}

One way to proceed with the study of back-reaction is to consider the expansion of regions of space in a given foliation. The equations that govern this expansion can then be found, and compared to the Friedmann equations. This often leads one to consider the volume-weighted average of quantities such as energy density and pressure. The equations that result are therefore often referred to as the `averaged field equations'.

While simple, this approach has a number of obvious drawbacks. Firstly, it is manifestly not foliation invariant. Secondly, there is a freedom in how one chooses to specify that two spatial volumes at different times are the same region. And thirdly, the averaging of quantities over the spatial volume being considered is often only well defined for scalars.  One can specify choices for the first and second of these that may initially appear natural, but that could in the end lead one to consider hyper-surfaces in the inhomogeneous space-time that become arbitrarily, and increasingly, distorted. The third of these problems is of more fundamental difficulty, as tensors cannot in general be compared at different points. Nevertheless, this approach provides a useful framework to investigate, and can be shown to give a straightforward correspondance to the average of observables in some situations.

\subsection{Buchert's Equations}

The most well studied set of averaged equations that result from this approach are those found by Buchert after averaging the Hamiltonian, Raychaudhuri and conservation equations \cite{buchert}:
\begin{align}
&3 \frac{\dot{a}^2_{\mathcal{D}}}{a_{\mathcal{D}}^2} = 8 \pi G_N \langle \rho \rangle_{\mathcal{D}} - \frac{1}{2} \left\langle^{(3)} R \right\rangle_{\mathcal{D}} - \frac{1}{2} \mathcal{Q}_{\mathcal{D}}
\label{buchert1}
\\
&3 \frac{\ddot{a}_{\mathcal{D}}}{a_{\mathcal{D}}} = - 4 \pi G_N \langle \rho \rangle_{\mathcal{D}} + \mathcal{Q}_{\mathcal{D}}
\label{buchert2}
\\
&\partial_t \langle \rho \rangle_{\mathcal{D}} + 3 \frac{\dot{a}_{\mathcal{D}}}{a_{\mathcal{D}}} \langle \rho \rangle_{\mathcal{D}} = 0,
\label{buchert3}
\end{align}
where $a_{\mathcal{D}}$ and $\langle^{(3)}R \rangle$ are the ``scale factor" and average Ricci curvature of the region of space $\mathcal{D}$ being considered, angular brackets denote a volume average throughout that region, and $\mathcal{Q}_{\mathcal{D}}$ is the back-reaction term that quantifies differences from the Friedmann equations that one might otherwise construct from these quantities. These are defined as
\begin{align}
&a_{\mathcal{D}}(t) = \left( \frac{\int_{\mathcal{D}} d^3 X \sqrt{\;^{(3)}g(t,X^i)}}{\int_{\mathcal{D}} d^3 X \sqrt{\;^{(3)}g(t_0,X^i)}} \right)^{\frac{1}{3}}
\\
&\left\langle \psi \right\rangle_{\mathcal{D}}(t) = \frac{\int_{\mathcal{D}} d^3 X \psi (t,X^i) \sqrt{\;^{(3)}g(t,X^i)}}{\int_{\mathcal{D}} d^3 X \sqrt{\;^{(3)}g(t,X^i)}}
\\
&\mathcal{Q}_{\mathcal{D}} = \frac{2}{3} \left( \left\langle \Theta^2 \right\rangle_{\mathcal{D}} - \left\langle \Theta \right\rangle^2_{\mathcal{D}} \right) - 2 \left\langle \sigma^2 \right\rangle_{\mathcal{D}},
\label{Q}
\end{align}
where $\Theta$ and $\sigma$ are the expansion and volume-preserving shear of the set of curves orthogonal to the hyper-surfaces containing $\mathcal{D}$, and $t$ and $X^i$ are the proper time measured along this set of curves and the spatial coordinates in the hyper-surfaces of constant $t$, respectively. The quantity $t_0$ is the value of $t$ on some reference hyper-surface (usually taken to be the one that contains us at present).

One may note that equations (\ref{buchert1})-(\ref{buchert3}) do not form a closed set. Extra information is therefore required, which can be given by specifying $\mathcal{Q}_{\mathcal{D}}= \mathcal{Q}_{\mathcal{D}}(t)$. Presumably this requires either extra equations, or some knowledge of the inhomogeneous space-time being averaged. As previously stated, one may also note that the averaging procedure given here by the angular brackets is foliation dependent and only applicable to scalars (this is particularly problematic for the term $\langle \sigma^2 \rangle_{\mathcal{D}}$ in equation (\ref{Q}), as the evolution equation for $\sigma^2$ will contain tensors). Finally, while the expansion of the spatial domain $\mathcal{D}$ may not itself be directly observable, we will explain below that in some cases it can be linked to observables.

\subsection{Links to Observables}

The term `observables' can cover a wide array of different possibilities in cosmology. Here we will mainly be concerned with the luminosity distance-redshift relation. This is itself a direct observable of considerable interest for the interpretation of, for example, supernova observations. Beyond this, it is also often required in the interpretation of other observables as it is very often the case that one needs to transform from ``redshift space" to some concept of position space (i.e. the position of astrophysical objects on some spatial hyper-surface).

The usual method for calculating luminosity distances in an inhomogeneous space-time is to first find the angular diameter distance to the emitting object as a function of some affine parameter, measuring distance along past-directed null geodesics. This can be achieved by integrating the Sachs optical equations\cite{sachs}. In these equations the Ricci curvature of the space-time sources the evolution of the expansion of the past-directed null geodesics, and the Weyl curvature sources the evolution of their volume-preserving shear (which itself acts as a source for their expansion). The angular-diameter distance can then be straightforwardly related to the luminosity distance\cite{etherington}, and the redshift can be calculated as a function of the affine distance (once the world-lines of the objects emitting the radiation have been specified). This then provides the luminosity distance as a function of redshift at all points on an observer's past-light cone, provided that geometric optics remains a good approximation, and that the light emitted from the distant object is not obscured by some intermediate matter before it reaches the observer.

Although the method outlined above is, in general, a complicated problem involving a number of subtleties, it was recently shown by R\"{a}s\"{a}nen\cite{rasanen} that progress can be made in space-times that display statistical homogeneity and isotropy on large scales. In this case one can estimate the average luminosity distance as a function of the average redshift that an observer in such a space-time may expect to reconstruct from observations made over cosmologically interesting distances. Assuming that the matter content is irrotational, that the shear in the null trajectories can safely be assumed to be small, that structures evolve slowly, and that hyper-surfaces of constant proper time can also be taken to be the same hyper-surfaces that display statistical homogeneity and isotropy, R\"{a}s\"{a}nen made a convincing case that the average luminosity distance-redshift relation in the inhomogeneous space-time should be well approximated by observables calculated in a homogeneous and isotropic model with a scale factor that evolves according to equations (\ref{buchert1})-(\ref{buchert3}).

An alternative approach to this problem was taken by Clarkson and Umeh\cite{clarkson}. These authors considered expressions for measures of distance expanded as a power series in redshift, as derived for general space-times by Kristian and Sachs\cite{ks}. They then performed a decomposition into spherical harmonics, and constructed the following deceleration parameter, based on an analogy between the monopole of this expansion and the corresponding relations in a Friedmann universe:
\begin{equation}
\label{cu}
q_0 = \frac{1}{H_0^2} \left[ \frac{4 \pi G}{3} \left( \rho + 3 p +12 \sigma^2 \right)\right]_0,
\end{equation}
where $H=\Theta/3$ is the isotropic part of the Hubble rate, and subscript `$0$' denotes a quantity evaluated at $z=0$. Using this expression they could consider the average deceleration within either a region of space, or a region of space-time. However, for matter obeying the strong-energy condition it can be seen from equation (\ref{cu}) that the average of $q_0$ will always be non-negative, and so the space-time (according to this measure) will always be inferred to be decelerating (in the absence of $\Lambda$). This is in contrast to the averaged evolutions possible from equations (\ref{buchert1})-(\ref{buchert3}), and at first glance would appear to contradict the results of R\"{a}s\"{a}nen described above.

In fact, there is no contradiction between these two sets of results\cite{bull}. That is, the observable calculated by Clarkson and Umeh should be expected to be a good approximation to the deceleration that one would infer from observations made within a small region around an observer. This measure is closely related to the acceleration of space within that region, as specified by Einstein's equations (as long as shear is small), and not by equations (\ref{buchert1})-(\ref{buchert3}). The observational measures considered by R\"{a}s\"{a}nen, however, are only expected to approach the evolution described by equations (\ref{buchert1})-(\ref{buchert3}) when the distances over which observations are made are much larger than the homogeneity scale of the space-time under consideration. This is, of course, the regime in which cosmological observations are usually made. Using example space-times it has been explicitly demonstrated that it is entirely possible for a set of observers in a given region of the universe to infer deceleration from Clarkson and Umeh's measure, while inferring acceleration from Buchert's measure\cite{bull}. This clearly demonstrates that the acceleration inferred from cosmological observations does {\it not} have to be closely related to the local acceleration of space itself. It also demonstrates that quantities that are uniquely defined in an exactly homogeneous and isotropic universe (such as $q_0$) can bifurcate into multiple different quantities in space-times that are only statistically homogeneous and isotropic, and that in general these new quantities can take very different values from each other. One must therefore proceed with care.

\section{Space-Time Averaging Approach}
\label{spacetime}

An alternative approach to considering the volume weighted average of quantities within 3-dimensional spatial regions is to instead consider averaging geometric quantities within 4-dimensional regions of space-time. Such a process is in general difficult to define in a covariant way, and so far has required the application of bi-local operators. These allow tensors to be compared at different points by transporting them along prescribed sets of curves. This then leads to the problems of how the curves in question should be prescribed, and exactly which transport method should be used. Various proposals exist as to the best way to address these issues\cite{rob}.

While complicated, the idea of averaging quantities in 4-dimensional regions of space-time inherently avoids any foliation dependence. These approaches are also often aimed at averaging tensors directly, rather than just scalars. This has obvious advantages for gravitational theories constructed from tensors, such as general relativity.

\subsection{Zalaletdinov's Equations}

Probably the most well known attempt at averaging in space-time, and constructing a set of effective field equations that the averages should obey, is that of Zalaletdinov\cite{zal}. The first step in this approach is to construct the following average for a tensor $ p^{\alpha \dots}_{\beta \dots}$:
\begin{equation}
\left\langle p^{\alpha \dots}_{\beta \dots} (x) \right\rangle = \frac{1}{V_{\Sigma}} \int_{\Sigma} \sqrt{-g^{\prime}} d^4 x^{\prime}  p^{\mu^{\prime} \dots}_{\nu^{\prime} \dots} (x^{\prime}) \mathcal{A}^{\alpha}_{\phantom{\alpha} \mu^{\prime}} (x,x^{\prime})  \mathcal{A}^{\nu^{\prime}}_{\phantom{\nu^{\prime}} \beta} (x,x^{\prime})  \dots ,
\end{equation}
where primed coordinates are those used in the 4-dimensional region $\Sigma$, which is the averaging domain associated with the point $x$. The quantities $ \mathcal{A}^{\alpha}_{\phantom{\alpha} \mu^{\prime}}$ are the bi-local operators, which are functions of both $x$ and $x^{\prime}$, and the quantity $V_{\Sigma}=\int_{\Sigma} \sqrt{-g^{\prime}} d^4 x^{\prime}$ is the volume of $\Sigma$. Each point, $x$,  is expected to have associated with it its own averaging domain, $\Sigma$, which is related to other averaging domains by being transported around the manifold.

By applying this averaging technique to the connection, and by using some ``splitting rules'', Zalaletdinov is able to use Einstein's equations to derive a set of field equations that the averaged connection must obey. The are called the Macroscopic Field Equations, and are written\cite{zal}
\begin{equation}
\label{MG}
\bar{g}^{\beta \epsilon} M_{\gamma \beta} -\frac{1}{2} \delta^{\epsilon}_{\phantom{\epsilon} \gamma} \bar{g}^{\mu \nu} M_{\mu \nu} = 8 \pi G \bar{T}^{\epsilon}_{\phantom{\epsilon} \gamma} - \left(Z^{\epsilon}_{\phantom{\epsilon} \mu \nu \gamma} - \frac{1}{2} \delta^{\epsilon}_{\phantom{\epsilon} \gamma} Q_{\mu \nu} \right) \bar{g}^{\mu \nu},
\end{equation}
where bars denote averaged quantities, and $M_{\gamma \beta} = M^{\alpha}_{\phantom{\alpha} \gamma \alpha \beta}$ and $Q_{\mu \nu} = Z^{\alpha}_{\phantom{\alpha} \mu \nu \alpha}$ and $Z^{\alpha}_{\phantom{\alpha} \mu \nu \beta} = 2 Z^{\alpha \phantom{\mu [ \epsilon} \epsilon}_{\phantom{\alpha} \mu [ \epsilon \phantom{\epsilon} \underline{\nu} \beta]}$, where
\begin{align}
&M^{\mu}_{\phantom{\mu} \nu \alpha \beta} = \partial_{\alpha} \langle \Gamma^{\mu}_{\phantom{\mu} \nu \beta} \rangle - \partial_{\beta} \langle \Gamma^{\mu}_{\phantom{\mu} \nu \alpha} \rangle + \langle  \Gamma^{\mu}_{\phantom{\mu} \sigma \alpha} \rangle \langle  \Gamma^{\sigma}_{\phantom{\sigma} \nu \beta} \rangle -  \langle  \Gamma^{\mu}_{\phantom{\mu} \sigma \beta} \rangle  \langle  \Gamma^{\sigma}_{\phantom{\sigma} \nu \alpha} \rangle
\\
&Z^{\alpha \phantom{\beta  \gamma} \mu}_{\phantom{\alpha} \beta \gamma \phantom{\mu} \nu \sigma}= \langle \Gamma^{\alpha}_{\phantom{\alpha} \beta [ \gamma} \Gamma^{\mu}_{\phantom{\mu} \underline{\nu} \sigma ]} \rangle - \langle \Gamma^{\alpha}_{\phantom{\alpha} \beta [ \gamma} \rangle \langle \Gamma^{\mu}_{\phantom{\mu} \underline{\nu} \sigma ]} \rangle ,
\end{align}
and where underlined indices are not included in symmetrization operations. The tensor $Z^{\alpha \phantom{\beta  \gamma} \mu}_{\phantom{\alpha} \beta \gamma \phantom{\mu} \nu \sigma}$ is known as the 2-point correlation tensor, and obeys its own algebraic and differential constraints\cite{zal}.

The Macroscopic Field Equations (\ref{MG}) can be used to describe the behavior of a particular inhomogeneous space-time after averaging has been performed, but they can also be used as a set of field equations to which one can look for solutions directly. This latter approach has so far been taken in the cases of macroscopic geometries, $\bar{g}_{\mu \nu}$, that are spatially homogeneous and isotropic\cite{coley,rob2}, and geometries that are spherically symmetric and static\cite{rob3}. This work has allowed some possible behaviors of averaged space-times to be found {\it without} specifying the underlying microscopic geometry. However, it has also so far required a number of assumptions to be made about the correlations that are present. These include the vanishing of the three-point and four-point correlation tensors, and the vanishing of the `electric' part of the 2-point correlation tensor\cite{rob2}. The particular situations in which these assumptions are valid remains to be determined, as is also the case for the assumptions that go into the derivation of the Macroscopic Field Equations (\ref{MG}). Nevertheless, this is an interesting approach that deserves further study.

\subsection{Links to Observables}

Under the assumption that the macroscopic geometry is spatially homogeneous and isotropic (that is, after the averaging procedure has been applied, and the ``averaged" geometry displays these symmetries), then Coley, Pelavas and Zalaletdinov find the following to be a solution of the Macroscopic Field Equations (\ref{MG})\cite{coley}:
\begin{equation}
\label{MGfrw}
\bar{g}_{\mu \nu} dx^{\mu} dx^{\nu} = -dt^2 + a^2(t) \left[ \frac{dr^2}{1-k_g r^2} + r^2 (d \theta^2 + \sin^2 \theta d \phi^2 ) \right],
\end{equation}
where $k_g$ is a constant, and where $a(t)$ and $\rho$ obeys the Friedmann-like equations
\begin{align}
&\frac{\dot{a}^2}{a^2}  = \frac{8 \pi G}{3} \rho - \frac{k_d}{a^2}
\label{MGfried}
\\
&\dot{\rho} + 3 \frac{\dot{a}}{a} (\rho +p) =0,
\label{MGdrho}
\end{align}
where $k_d$ is a constant (not necessarily equal to $k_g$), and where $\rho$ and $p$ are the macroscopic energy density and pressure (obtained after averaging the right-hand side of Einstein's equations).

Superficially, the geometry given in equations (\ref{MGfrw})-(\ref{MGdrho}) looks a lot like the spatially homogeneous and isotropic solutions of Einstein's equations in the presence of a perfect fluid. There is, however, a very significant difference: The spatial curvature constant that appears in the macroscopic geometry, $k_g$, is not in general the same as the term that looks like spatial curvature in the Friedmann-like equation (\ref{MGfried}) (i.e. the one that contains $k_d$). That is, spatially curvature can take different values depending on the situation being considered. If one measured the angles at the corners of triangle, and determined the curvature of space in this way, then this would give a different result to that which would be obtained by measuring the recessional velocities of astrophysical objects and inferring the spatial curvature through the dynamical (Friedmann-like) equation (\ref{MGfried}). This behavior is impossible within the spatially homogeneous and isotropic solutions of Einstein's equations, and so could provide some potentially observable phenomena that could be used to test this approach. The difference between $k_g$ and $k_d$ is determined by terms that appear in the correlation tensor, $Z^{\alpha \phantom{\beta  \gamma} \mu}_{\phantom{\alpha} \beta \gamma \phantom{\mu} \nu \sigma}$, and so by attempting to determine the difference between $k_g$ and $k_d$ observationally we could attempt to constrain $Z^{\alpha \phantom{\beta  \gamma} \mu}_{\phantom{\alpha} \beta \gamma \phantom{\mu} \nu \sigma}$, and hence some of the possible effects of averaging.

A first step towards investigating this possibility has recently been taken\cite{sung}. The authors of this work assume that average observables are determined by null trajectories in the average geometry, as specified in equation (\ref{MGfrw}), and that redshifts are represented by the average scale factor, $a(t)$. They then find that luminosity distances are given by the following equation:
\begin{equation}
\label{MGdl}
d_L(z) = \frac{(1+z)}{H_0 \sqrt{\vert \Omega_{k_g} \vert}} f_{k_g} \left( \int_{\frac{1}{1+z}}^1 \frac{\sqrt{\vert \Omega_{k_g} \vert} da} {\sqrt{ \Omega_{k_d} a^2 + \Omega_{\Lambda} a^4 + \Omega_m a}} \right),
\end{equation}
where the matter content of the macroscopic space-time has been assumed to be well approximated by non-interacting dust and $\Lambda$, where $H_0=\dot{a}/a\vert_{z=0}$, and where the $\Omega_i$ are defined as
\begin{equation}
\Omega_{k_g} =  -\frac{k_g}{a_0^2 H_0^2}, 
\hspace{0.5cm} 
\Omega_{k_d} = -\frac{k_d}{a_0^2 H_0^2} ,
\hspace{0.5cm} 
\Omega_{m} = \frac{8 \pi G \rho_{m,0}}{3 H_0^2},
\hspace{0.5cm} 
\Omega_{\Lambda} = \frac{8 \pi G \rho_{\Lambda}}{3 H_0^2},
\end{equation}
where $\rho_{m,0}$ is the present energy density in dust, and $8 \pi G \rho_{\Lambda} = \Lambda$ is the effective energy density in $\Lambda$. The expression for luminosity distance given in equation (\ref{MGdl}) can now be used to interpret cosmological observations, and to obtain constraints on the $\Omega_i$.

Using data from the Hubble Space Telescope (HST)\cite{hst}, the Wilkinson Microwave Anisotropy Probe (WMAP)\cite{wmap}, observations of the Baryon Acoustic Oscillations (BAOs)\cite{bao}, and the Union2\cite{union2} and SDSS\cite{sdss} supernova data sets, the parameters $\Omega_{k_d}$, $\Omega_{k_g}$ and $\Omega_{\Lambda}$ were constrained to take the values given in Table \ref{table1} below\cite{sung}. The additional freedom of allowing $\Omega_{k_d} \neq \Omega_{k_g}$ in this analysis means that the CMB+$H_0$ is now no longer sufficient to constrain the spatial curvature of the universe significantly. Observations of the CMB+$H_0$ alone are also no longer sufficient to require $\Lambda \neq 0$. This simple extra degree of freedom therefore undermines two of the most important results of modern observational cosmology. By adding further data sets the constraints on $\Omega_{k_d}$ and $\Omega_{k_g}$ are improved, but still remain much weaker than in the standard Friedmann models that satisfy Einstein's equations. Even so, however, it was still found that the results of using all available observables were sufficient to require $\Omega_{\Lambda} \neq 0$ to high confidence, and that a spatially flat universe was consistent with observations. Finally, although the combination of {\it some} data sets excluded the possibility $\Omega_{k_d}=\Omega_{k_g}$ at the 95\% confidence level, it was found that the special case $\Omega_{k_d}=\Omega_{k_g}$ was compatible with {\it most} combinations of these data sets.

\begin{table}[ht!]
\tbl{Constraints on $\Omega_{k_d}$, $\Omega_{k_g}$ and $\Omega_{\Lambda}$ from data sets outlined in the text.}
{\begin{tabular}{c|ccc}
\hline \hline
 Data Sets  &  $\Omega_{k_d}$ & $\Omega_{k_g}$ & $\Omega_{\Lambda}$    
\\  [0.5mm] 
\hline
CMB & $-0.053^{+0.152}_{-0.153}$ & $-0.036^{+0.562}_{-0.572}$ & $+0.525^{+0.417}_{-0.524}$
\\
CMB+HST & $+0.036^{+0.062}_{-0.064}$ & $+0.185^{+0.396}_{-0.415}$ & $+0.564^{+0.415}_{-0.401}$
\\
SNIa (Union2) & $+0.012^{+0.513}_{-0.485}$ & $-0.369^{+0.398}_{-0.410}$ & $+0.902^{+0.189}_{-0.187}$
\\
SNIa (SDSS) & $+0.233^{+0.466}_{-0.451}$ & $-0.173^{+0.492}_{-0.507}$ & $+0.641^{+0.230}_{-0.225}$
\\
CMB+HST+SNIa(Union2) & $+0.014^{+0.017}_{-0.017}$ & $+0.055^{+0.092}_{-0.092}$ & $+0.695^{+0.080}_{-0.082}$ 
\\
CMB+HST+SNIa(SDSS) & $+0.054^{+0.020}_{-0.020}$ & $+0.311^{+0.100}_{-0.101}$ & $+0.436^{+0.087}_{-0.089}$
\\
CMB+HST+SNIa(Union2)+BAO & $-0.004^{+0.011}_{-0.011}$ & $-0.033^{+0.070}_{-0.069}$ & $+0.755^{+0.068}_{-0.070}$ 
\\
CMB+HST+SNIa(SDSS)+BAO & $+0.026^{+0.012}_{-0.012}$ & $+0.183^{+0.072}_{-0.070}$ & $+0.522^{+0.070}_{-0.073}$ 
\\
\hline \hline
\end{tabular}
\label{table1}}
\end{table}

In general one might also consider the possibility of not just allowing $\Omega_{k_d}$ and $\Omega_{k_g}$ to be different, but also allowing them to functions of scale. Such a result might arise, for example, from performing averaging over domains of different sizes, a process which is implicitly carried out when consider different cosmological observables. Such a possibility allows for considerable extra freedom\cite{sung}.

\section{Constructing Inhomogeneous Models}
\label{models}

We have so far considered attempts to describe the large-scale behavior of the universe by averaging over regions of space or space-time. In the end, the particular approach that one should use when performing this type of operation should probably be guided by the phenomena that one is trying to create a model to interpret. Different observable phenomena may require different approaches, and so one needs to know the limits of any particular approach, as well as the situations in which it reliably reproduces the required results. For this it is useful to have inhomogeneous cosmological models that are of sufficient generality to allow some of the interesting behavior that is expected in general. Such models can then be used to test ideas about averaging, back-reaction, and the large-scale evolution of space.

Unfortunately it is extremely difficult to construct such models. This does {\it not} mean that there is an absence of any interesting behavior to study, only that we need to become more sophisticated in our model building to quantify and constrain the different possibilities in a reliable way. Some of the principal difficulties involved with this are how to model over-dense regions of the universe without having to deal with the rapid formation of singularities, how to introduce structure into the universe without assuming a Friedmann background or matching onto a Friedmann model at a boundary, and how to allow structure to form on different scales without assuming linearity in the field equations. For further discussion of inhomogeneous cosmological solutions the reader is referred to the contribution to these proceedings by Krasi\'{n}ski \cite{kra2}, and to the comprehensive texts \cite{kra,kra3}.

It is currently almost beyond hope to construct a model that allows for all of the possibilities discussed above, while simultaneously maintaining sufficiently generality to model realistic distributions of matter. We are therefore forced to investigate toy models that we hope may reflect some of the features of the real universe, even if they are not realistic in every way. Once toy models have been constructed we can then consider the averaging problem by applying some of the methods discussed above to them, or by fitting or comparing them to Friedmann models directly. Their existence also makes more advanced models a more realistic proposition. It is for these reasons that it is of interest to consider simple $n$-body solutions of Einstein's equations. Such solutions, if they can be found, will allow over-dense regions to be studied without rapid collapse occurring, and without recourse to the assumption of a Friedmann background or linearity in the gravitational field equations. This will be the subject of Section \ref{bhsec}.

\subsection{A Lattice of Black Holes}
\label{bhsec}

The simplest configuration of $n$ bodies that one can imagine is a regularly arranged set of points. Although such a configuration limits the behaviors that are possible, it does allow for the most straightforward possible comparison to smoothed-out Friedmann-like universes. That is, by `zooming out' in order to consider large numbers of points, and by performing some kind of coarse graining or smoothing, one could easily imagine such a situation looking more and more like a spatially homogeneous and isotropic universe, which could then be compared to the Friedmann solutions of Einstein's equations. Regularity of the distribution also provides a limited number of preferred spatial planes and curves that can have their area and length compared to those of the Friedmann solutions.

Here we will consider spatially closed universes. These are known to admit hyper-surfaces of time symmetry at the maximum of expansion of the space-time that allow the constraint equations to be solved in a particularly simple way\cite{ts}. The method that we will deploy to ensure that our massive bodies are regularly arranged is to tile the hyper-surface of maximum expansion with a number of regular polyhedra. A mass is then placed at the center of each polyhedron, which by symmetry will be an equal distance from each of its nearest neighbors. These polyhedra will be referred to in what follows as `cells'. There are seven such tilings that are possible in three spatial dimensions, as listed in Table \ref{table2}. Also displayed in this table are the Schl\"{a}fli symbols of the polychora that these tilings constitute\cite{schlafli}.

\begin{table}[ht!]
\tbl{Tilings of the 3-space of maximum expansion, and their scale in comparison to the homogeneous and isotropic Friedmann solutions.}
{\begin{tabular}{c|c|c|c}
\hline \hline
Lattice & Cell & Number of & Ratio of scales in discrete
\\
Structure & Shape & Cells & and Friedmann solutions\cite{kjell}
\\  [0.5mm] 
\hline
- & Ball & $2$ & -
\\
\{333\} & Tetrahedron & $5$ & 1.360
\\
\{433\} & Cube & $8$ & 1.291
\\
\{334\} & Tetrahedron & $16$ & 1.097
\\
\{343\} & Octahedron & $24$ & 1.099	
\\
\{533\} & Dodecahedron & $120$ & 1.034
\\
\{335\} & Tetrahedron & $600$ & 1.002
\\
\hline \hline
\end{tabular}
\label{table2}}
\end{table}

\begin{figure}
\centering
  \subfloat[A slice through the 8-cell.]{\label{8fig}
    \includegraphics[height=5.5cm]{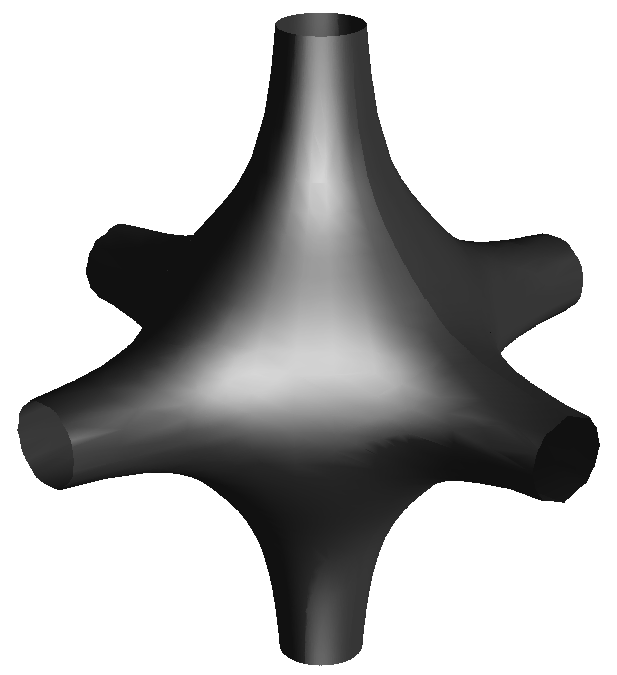}}
\hspace{1cm}
  \subfloat[A slice through the 120-cell.]{\label{120fig}
    \includegraphics[height=5.5cm]{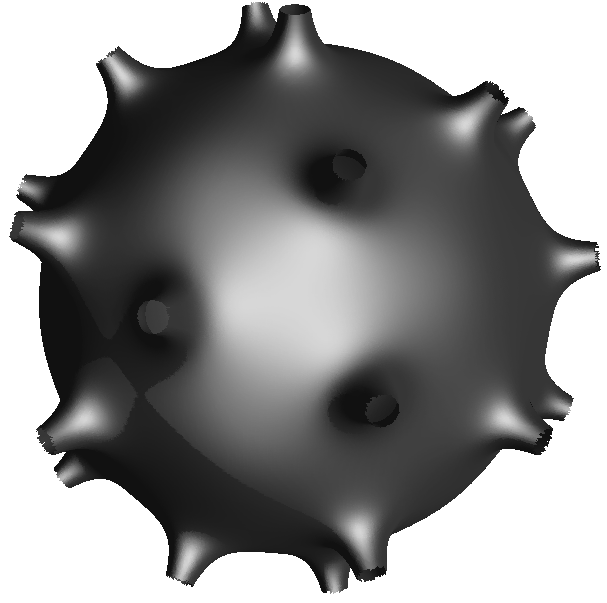}}
\vspace*{8pt}
\caption{Graphic illustrations of the geometry of space at the maximum of expansion. \label{f1}}
\end{figure}

Once the arrangement of masses has been chosen, the geometry of the hyper-surface of maximum-of-expansion can be found. With the exception of the 2-cell, this has been done for each of the structures described above\cite{kjell}. The 2-cell is special in that the time-symmetric geometry at the maximum of expansion of this structure is simply a slice through the global Schwarzschild solution. The geometry of the full space-time is therefore already known exactly in this case, and is not of cosmological interest here (by including a non-zero $\Lambda$, however, other structures are also possible\cite{uzan}). An illustration of the geometry at the maximum of expansion in the case of the 8-cell and the 120-cell is given in Figure \ref{f1}, below. Each of the illustrations here corresponds to a single 2-dimensional slice through the 3-dimensional geometry. In the case of the 8-cell this slice contains 6 masses, while in the 120-cell it contains many more (although not all 120). The geometry of the 3-space of maximum of expansion in each case is conformally related to the geometry of a 3-sphere, with a scale factor that is a function of position. The distance from the origin in the illustrations in Figure \ref{f1} is proportional to this scale factor, and it can be seen that as the number of the masses in the lattice is increased, the bulk of the space approaches homogeneity. It is only in the vicinity of the masses themselves that inhomogeneities exist (as depicted by the tube-like structures).

Once we have the geometry of the hyper-surface of maximum of expansion, we can take a measure of the scale of the solution, and compare this to the scale of a spatially closed Friedmann universe that contains the same amount of ``proper mass\cite{kjell}''. The Friedmann solutions will, of course, have this mass evenly distributed throughout space, and so by comparing to the scale of the inhomogeneous geometry we can obtain a measure of back-reaction.  For the choice of scale in the inhomogeneous space on could choose a number of different measures. Here we consider the proper length of the edge of a cell. This corresponds to the scale of curvature for the sphere that appears to emerge when the number of masses becomes large (as can be seen from the illustration in Figure \ref{120fig}). The difference in scale in each case is given in the last column of Table \ref{table2}. It can be seen that the broad trend is for the scale of the homogeneous and inhomogeneous space-times to approach each other as the number of cells becomes large. However, for only a small number of cells ($\sim 5$ to $24$) the difference in scale can be of order $10\%$. In any case, this method provides an exact quantification of back-reaction, and provides an arena for testing formalisms designed for more general configurations of energy and momentum.

A numerical evolution of the 8-cell has now been performed\cite{bent}, and other methods have also been used to address problem of understand the evolution of this type of structure\cite{evo1,evo2,evo3}.

\section{Discussion and Outlook}
\label{discussion}

Various approaches to back-reaction and averaging already exist in the literature, but much work remains to be done if we are to fully understand their observational consequences in the real universe. Motivation for taking these problems seriously comes from the apparent necessity of including dark energy when we interpret observations within a linearly perturbed Friedmann model, as well as the requirement to understand all possible sources of error and uncertainty in precision cosmology. To fully address this problem it is likely that we will need to develop more sophisticated models of inhomogeneous space-times, as well as developing a more sophisticated understanding of averaging in general relativity. Research in this area should be considered exceptionally timely, with large amounts of resources currently being invested into observational probes designed to improve our understanding of dark energy, and the universe around us.

\section*{Acknowledgments}

I acknowledge the support of the STFC.

%\section{References}

\end{document}